\documentstyle[aps,epsfig]{revtex}

\def\fun#1#2{\lower3.6pt\vbox{\baselineskip0pt\lineskip.9pt
\ialign{$\mathsurround=0pt#1\hfil##\hfil$\crcr#2\crcr\sim\crcr}}}

\begin{document}

\title{Determination of the quarkonium--gluonium content of the
isoscalar tensor resonances $f_2(1920)$, $f_2(2020)$, $f_2(2240)$,
$f_2(2300)$ and of the broad state $f_2(2000)$ based on decay couplings
to $\pi^0\pi^0,\eta\eta,\eta\eta'$}

\author{V.V. Anisovich$^a$, M.A. Matveev$^a$, J. Nyiri$^b$ and
A.V.~Sarantsev$^{a,c}$\\
$^a$ Petersburg Nuclear Physics Institute, Gatchina 188300, Russia\\
$^b$ KFKI Research Institut for Particle and Nuclear Physics, Budapest,
Hungary\\
$^c$ HISKP, Universit\"at Bonn, D-53115 Germany}

\date{01.01.2005}

\maketitle

\begin{abstract}
In the reactions $p\bar p\to\pi^0\pi^0,\eta\eta,\eta\eta'$ there are
four relatively narrow resonances $f_2(1920)$, $f_2(2020)$, $f_2(2240)$,
$f_2(2300)$, and a broad one $f_2(2000)$ in the mass region
1990--2400~MeV. In the framework of quark combinatorics we carry out
an analysis of the decay constants for all five resonances. It is shown
that the relations for the decay constants corresponding to the broad
resonance $f_2(2000)\to\pi^0\pi^0,\eta\eta, \eta\eta'$ are the same as
those corresponding to a glueball. An additional argument in favour of
the glueball--nature of $f_2(2000)$ is the fact that $f_2(1920)$,
$f_2(2020)$, $f_2(2240)$, $f_2(2300)$ fit well the $q\bar q$
trajectories in the $(n,M^2)$-plane (where $n$ is the radial quantum
number), while the broad $f_2(2000)$ resonance turns out to be an
unnecessary extra state for these trajectories.
\end{abstract}


\section{Introduction}

A broad $f_2$-resonance was observed in the region of 2000 MeV in various
reactions (see the compilation \cite{PDG} and references therein).
The measured masses and widths are: \\
$M=2010\pm25\,$MeV, $\Gamma=495\pm35\,$MeV in $p\bar p\to\pi^0\pi^0,
\eta\eta, \eta\eta'$ \cite{Ani},\\
$M=1980\pm20\,$MeV, $\Gamma=520\pm50\,$MeV in $pp\to pp4\pi$ \cite{Bar},\\
$M=2050\pm30\,$MeV, $\Gamma=570\pm70\,$MeV in $\pi^-p\to\phi\phi n$
\cite{LL}.\\
Following these measurements, we denote the broad resonance as
$f_2(2000)$.

A recent re-analysis of the $\phi\phi$ spectra \cite{LL} in the reaction
 $\pi^-p\to\phi\phi n$ \cite{Etk}, and the analysis of the process
$\gamma\gamma\to K_SK_S$ \cite{L3} has essentially clarified
the status of the $(I=0$, $J^{PC}=2^{++}$)-mesons. This enables us to
place them on the $(n,M^2)$-trajectories, where $n$ is the radial quantum
number of the $q\bar q$-state.

In \cite{syst} (see also \cite{book,ufn04,glueball2}) we put the known
$q\bar q$-mesons consisting of light quarks ($q=u,d,s$) on the $(n,M^2)$
trajectories. Trajectories for mesons with various quantum numbers turn
out to be linear with a good accuracy:
\begin{equation}
\label{1}
M^2\ =\ M^2_0+(n-1)\mu^2
\end{equation}
where $\mu^2=1.2\pm0.1\,\rm GeV^2$ is a universal slope, and $M_0$ is the
mass of the lowest state with $n=1$.

On Fig. 1 we show the present status of the $(n,M^2)$ trajectories for
the $f_2$-mesons (i.e. we use the results given by \cite{LL,L3}). To
avoid confusion, we list here the experimentally observable masses.

In \cite{LL} the $\phi\phi$ spectra are re-analysed, taking into account
the existence of the broad $f_2(2000)$ resonance. As a result, the
masses of three relatively narrow resonances are shifted compared to those
given in the compilation  PDG \cite{PDG}:

\begin{eqnarray}
&& f_2(2010)|_{PDG}\ \to\, f_2(2120)\ \cite{LL}\ ,
\nonumber\\
&& f_2(2300)|_{PDG}\ \to\, f_2(2340)\ \cite{LL}\ ,
\nonumber\\
&& f_2(2340)|_{PDG}\ \to\, f_2(2410)\ \cite{LL}\ .
\label{Int1}
\end{eqnarray}
A phase analysis for the reaction $\gamma\gamma\to K_SK_S$ is carried
out in \cite{L3}; the characteristics of $f_2(1755)$ are measured. This
resonance belongs to the nonet of the first radial excitation, $(n=2)$;
it is dominated by a $^3P_2\,s\bar s$ state. Its partner in the $(n=2)$
nonet is $f_2(1580)$. Neglecting a possible admixture of a glueball
component, in \cite{L3}
\begin{eqnarray}
f_2(1580) &=& n\bar n\cos\varphi_{n=2}+s\bar s\sin\varphi_{n=2}\ ,
\nonumber\\
f_2(1755) &=& -n\bar n\sin\varphi_{n=2}+s\bar s\cos\varphi_{n=2}\ ,
\nonumber\\
\label{Int2}
&&\varphi_{n=2}\ =\
-10^\circ\,{}^{\displaystyle+5^\circ}_{\displaystyle-10^\circ}
\end{eqnarray}
is found. Here $n\bar n=(u\bar u+s\bar s)/\sqrt2$.

The quark states with ($I=0$, $J^{PC}=2^{++}$) are determined by two
flavour components $n\bar n$ and $s\bar s$ for which two states
$^{2S+1}L_J=\,^3P_2,\,^3F_2$ are possible. Consequently, we have four
trajectories on the $(n,M^2)$ plane. Generally speaking, the $f_2$-states
are mixtures of both the flavour components and the $L=1,3$ waves. The
real situation is, however, such that the lowest trajectory
[$f_2(1275)$, $f_2(1580)$, $f_2(1920)$, $f_2(2240)$] consists of mesons
with dominant $^3P_2n\bar n$ components, while the trajectory
$[f_2(1525),f_2(1755),f_2(2120),f_2(2410)]$ contains mesons with
predominantly $^3P_2s\bar s$ components, and the $F$-trajectories are
represented by three resonances [$f_2(2020)$,$f_2(2300)$] and
[$f_2(2340)$] with the corresponding dominant $^3F_2n\bar{n}$ and $^3F_2s
\bar s$ states. Similarly to \cite{glueball2}, the broad resonance
$f_2(2000)$ is not part of those states placed on the $(n,M^2)$
trajectories. In the region of 2000~MeV we see three resonances,
$f_2(1920)$, $f_2(2000)$, $f_2(2020)$, while on the $(n,M^2)$-trajectories
there are only two vacant places. This means that one state is obviously
"superfluous" from the point of view of the $q\bar q$-systematics, i.e.
it has to be considered as exotics. The large value of the width of
$f_2(2000)$ strengthen the suspicion that, indeed, this state is an
exotic one.

In \cite{PR-exotic} it was pointed out that an exotic state has to
be broad. Indeed, if an exotic resonance occurs among the usual
$q\bar q$-states, they overlap, and their mixing becomes possible owing
to the $resonance\,(1)\to real\ mesons \to resonance\,(2)$ transitions
at large distances. It is due to these transitions that an exotic meson
accumulates the widths of its neighbouring resonances. The phenomenon
of the accumulation of widths was shown in the scalar sector near 1500~MeV
\cite{APS-PL,AAS-PL}. In fact the accumulation of widths was observed
much earlier, in overlapping resonances in nuclear physics \cite{shapiro,
okun,stodolsky}. Hence, the large width of $f_2(2000)$ can indicate that
this state is an exotic one. Still, to verify that $f_2(2000)$ is of
glueball nature, it is necessary to investigate the relations between
the coupling constants of $f_2(2000)$ with the meson channels. The
coupling constants of the transitions
$$
f_2(1920), f_2(2000), f_2(2020), f_2(2240),f_2(2300)\to\pi\pi^0,
\eta\eta,\eta\eta'
$$
were obtained in \cite{Ani,PNPI} from the analysis of the
reactions $p\bar p\to\pi^0\pi^0,\eta\eta,\eta\eta'$. To investigate the
quarkonium $(q\bar q)$ and gluonium $(gg)$ contents of $f_2(1920)$,
$f_2(2000)$, $f_2(2020)$, $f_2(2240)$, $f_2(2300)$, we make use of these
coupling constants in Section~2. It turns out that only the coupling
constants corresponding to $f_0(2000) \to\pi^0\pi^0,\eta\eta,\eta\eta'$
satisfy the relations which characterize the glueball (they are close to
those given by a flavour singlet). In the Conclusions we discuss the
flavour content of the $q\bar{q}$ states, and compare the properties of a
$2^{++}$ glueball with those of a $0^{++}$ glueball, situated in the
1500~MeV region.

\section{The determination of the 
$q\bar q-gg$ content of isoscalar tensor mesons, observed in the
reactions $p\bar p\to\pi^0\pi^0, \eta\eta,\eta\eta'$}

On the basis of data given by the analysis of the reactions
$p\bar p\to\pi^0\pi^0,\eta\eta,\eta\eta'$ \cite{Ani}, in this section we
investigate the quarkonium--gluonium content of $f_2(1920)$, $f_2(2000)$,
$f_2(2020)$, $f_2(2240)$, $f_2(2300)$. We do this in terms of the
rules of quark combinatorics.

\subsection{The rules of quark combinatorics for decay constants}

The isoscalar tensor $q\bar q$ mesons, close to the $2^{++}$ gluonium
state and mixing with it, are superpositions of the components
$\sqrt{1-W}\,q\bar q+\sqrt W\,gg$. The quarkonium component is, in its
turn, a mixture of $n\bar n=(u\bar u+d\bar d)/\sqrt2$ and $s\bar s$:
\begin{equation}
\label{4}
q\bar q\ =\ n\bar n\cos\varphi+s\bar s\sin\varphi\ .
\end{equation}
In terms of the $1/N$ expansion rules \cite{t'hooft} the transitions of
the $q\bar q$ and $gg$ components are not suppressed (the contribution of
the quark loop in the gluon ladder is of the order of $N_f/N_c$, where
$N_f$ is the number of light flavours, $N_c$ that of the colours; see,
e.g., Subsection 5.4.4 in \cite{book} and references therein). Hence,
it is justified to expect relevant admixtures of gluonic components to
states which we consider $q\bar q$-mesons (i.e. to those lying on the
$(n,M^2)$ trajectories in Fig.~1).

On the other hand, a glueball state has to contain essential $q\bar q$
components:
\begin{eqnarray}
\label{5a}
&& gg\cos\gamma+(q\bar q)_{glueball}\sin\gamma\ ,
\nonumber\\
&& (q\bar q)_{glueball}\ =\ n\bar n\cos\varphi_{glueball}+s\bar s
\sin\varphi_{glueball}\ .
\end{eqnarray}
If the flavour SU(3) symmetry were satisfied, the quarkonium component
$(q\bar q)_{glueball}$ would be a flavour singlet. In reality, the
probability of strange quark production in a gluon field is suppressed:
$u\bar u:d\bar d:s\bar s=1:1:\lambda$, where $\lambda\simeq0.5-0.85$.
Hence, $(q\bar q)_{glueball}$ differs slightly from the flavour singlet:
$(q\bar q)_{glueball}=(u\bar u+d\bar d+\sqrt\lambda\,s\bar s)/\sqrt{2+
\lambda}$ \cite{Alexei}. The suppression parameter $(\lambda)$ for the
production of a strange quark was estimated both in multiple hadron
production processes \cite{lambda}, and in hadronic decay processes
\cite{klempt,kmat}. According to these estimations, $\lambda$ is of the
order of 0.5--0.85, which leads to
\begin{equation}
\label{6a}
\varphi_{glueball}\ \simeq\ 26^0-33^0.
\end{equation}
The hadronic decay processes of quarkonium and gluonium states are
determined by production processes of new quark-antiquark pairs. In the
leading terms of the $1/N$ expansion the vertices of hadronic resonance
decays are determined by planar diagrams. Examples of planar diagrams for
the decay of a $q\bar q$ state and of a glueball into two $q\bar q$
mesons are shown in Fig.~2; in the decay process of a $q\bar q$ state
the gluons produce a new $q\bar q$-pair, while the decay of a glueball
leads to the production of two $q\bar q$-pairs.

For the $f_2\to\pi^0\pi^0,\eta\eta,\eta\eta'$ transitions, when the $f_2$
states are mixtures of quarkonium and gluonium components, the rules
of quark combinatorics give the following relations for the vertices
determined by planar diagrams \cite{book,kmat}:
\begin{eqnarray}
&& g_{\pi^0\pi^0}\ =\ g\,\frac{\cos\varphi}{\sqrt2}+\frac
G{\sqrt{2+\lambda}}\ ,
\nonumber\\
&&
g_{\eta\eta}=g\left(\cos^2\theta\frac{\cos\varphi}{\sqrt2}+\sin^2\Theta
\sqrt\lambda \sin\varphi\right)+\frac G{\sqrt{2+\lambda}}
(\cos^2\Theta+\lambda\sin^2\Theta)\ ,
\nonumber\\
&& g_{\eta\eta'}=\sin\Theta\cos\Theta\left[g\left(
\frac{\cos\varphi}{\sqrt2}-\sqrt\lambda\sin\varphi\right)+
\frac G{\sqrt{2+\lambda}}(1-\lambda)\right].
\label{5}
\end{eqnarray}
The terms proportional to $g$ stand for the $q\bar q\to two\,mesons$
transitions, while the terms with $G$ represent $glueball\to two\,mesons$.
Consequently, $g^2$ and $G^2$ are proportional to the probabilities for
finding quark-antiquark $(1-W)$ and glueball $(W)$ components in the
considered $f_2$-meson: $g^2=(1-W)g^2_0$ and $G^2=WG^2_0$, where $g_0$
and $G_0$ are couplings for pure states i.e. for transitions of the
type presented in Fig.~2.

Let us consider decays into a definite channel, for example
$q\bar q\to\pi^0\pi^0$ and $gg\to\pi^0\pi^0$. If so, the flavour
content of the quark loop is fixed, and the rules of the $1/N$ expansion
give $g(q\bar q\to\pi^0\pi^0)\sim1/\sqrt{N_c}\,$, $\,g(gg\to\pi^0\pi^0)
\sim 1/N_c$. Normalizing the constants as in (\ref{5}), this means
$G^2_0/g^2_0\sim1/N_c$. $\Theta$ is the mixing angle for the $n\bar n$
and $s\bar s$ components in the $\eta$ and $\eta'$ mesons ($\eta=n\bar n
\cos\Theta-s\bar s\sin\Theta$ and $\eta'=n\bar n\sin\Theta+s\bar s\cos
\Theta$); here we neglect the possible admixture of a glueball component
to $\eta$ and $\eta'$ (according to \cite{eta-glue}, the glueball
admixture to $\eta$ is less than 5\%, to $\eta'$ --- less than 20\%).
For the mixing angle $\Theta$ we assume $\Theta=37^\circ$.

Considering a glueball state, we have $\varphi\to\varphi_{glueball}$; if
so, the relations (\ref{5a}) turn into
\begin{eqnarray}
&& g_{\pi^0\pi^0}\ \to\, g^{(glueball)}_{\pi^0\pi^0}\ =\
\frac{g+G}{\sqrt{2+\lambda}}\ ,
\nonumber\\
&& g_{\eta\eta}\ \to\, g^{(glueball)}_{\eta\eta}\ =\
\frac{g+G}{\sqrt{2+\lambda}}\,(\cos^2\Theta+\lambda\sin^2\Theta)
\nonumber\\
&& g_{\eta\eta'}\ \to\, g^{(glueball)}_{\eta\eta'}\ =\
\frac{g+G}{\sqrt{2+\lambda}}\,(1-\lambda)\sin\Theta\cos\Theta\ .
\label{6}
\end{eqnarray}
Hence, in spite of the unknown quarkonium components in the glueball,
there are definite relations between the couplings of the glueball state
with the channels $\pi^0\pi^0,\eta\eta,\eta\eta'$ which can serve as
signatures to define it.

\subsection{Data for relations between coupling constants of 
$f_2$ resonance decays into $\pi^0\pi^0,\eta\eta,\eta\eta'$
channels indicating that $f_2(2000)$ is a glueball}

The analysis of the reactions $p\bar p\to \pi^0\pi^0,\eta\eta,\eta\eta'$
carried out in \cite{Ani,PNPI} provides us with the
parameters of the tensor resonances
$f_2(1920)$, $f_2(2000)$, $f_2(2020)$, $f_2(2240)$, $f_2(2300)$. In
Fig. 3, we demonstrate the cross sections for
$p\bar p\to \pi^0\pi^0,\eta\eta,\eta\eta'$ in $^3P_2$ and $^3F_2$ waves
(dashed and dotted lines) and the total $(J=2)$ cross section (solid
line) as well as Argand-plots for the $^3P_2$ and $^3F_2$ wave
amplitudes at invariant masses
$M=1.962$, $2.050$, $2.100$, $2.150$, $2.200$, $2.260$, $2.304$,
$2.360$, $2.410$ GeV.
 The data for $p\bar p\to \pi^+\pi^-$ \cite{Ei,Ca}
at $1.900\le M\le 2.200$ GeV are also included into  analysis.

 These amplitudes give us the following ratios
$g_{\pi^0\pi^0}:g_{\eta\eta}:g_{\eta\eta'}$ for the $f_2$ mesons:
\begin{eqnarray}
f_2(1920)\hspace*{2cm} &&\ 1:0.56\pm0.08:0.41\pm0.07
\nonumber\\
f_2(2000)\hspace*{2cm} &&\ 1:0.82\pm0.09:0.37\pm0.22
\nonumber\\
f_2(2020)\hspace*{2cm} &&\ 1:0.70\pm0.08:0.54\pm0.18
\nonumber\\
f_2(2240)\hspace*{2cm} &&\ 1:0.66\pm0.09:0.40\pm0.14
\nonumber\\
f_2(2300)\hspace*{2cm} &&\ 1:0.59\pm0.09:0.56\pm0.17.
\label{7}
\end{eqnarray}
For the glueball state the relations between the coupling constants are
$1:(\cos^2\Theta+\lambda\sin^2\Theta):(1-\lambda)\cos\Theta\sin\Theta$.
For $(\lambda=0.5$, $\Theta=37^\circ)$ we have $1:0.82:0.24$, and for
$(\lambda=0.85$, $\Theta=37^\circ)$, correspondingly, $1:0.95:0.07$.
Consequently, the relations between the coupling constants
$g_{\pi^0\pi^0}:g_{\eta\eta}:g_{\eta\eta'}$ for the glueball have to be
\begin{equation}
\hspace*{-3cm} 2^{++}glueball \hspace{1cm}
g_{\pi^0\pi^0}:g_{\eta\eta}:g_{\eta\eta'}\ =\
1:(0.82-0.95):(0.24-0.07).
\label{8}
\end{equation}
It follows from the expression (\ref{7}) that only the coupling constants
of the broad $f_2(2000)$ resonance are inside the $0.82\le g_{\eta\eta}/
g_{\pi^0\pi^0}\le0.95$ and $0.24\ge g_{\eta\eta'}/g_{\pi^0\pi^0}\ge0.07$
intervals. Hence, it is just this resonance which can be considered as a
candidate for a tensor glueball, while $\lambda$ is fixed in the interval
$0.5\le\lambda\le0.7$. Taking into account that there is no place for
$f_2(2000)$ on the $(n,M^2)$-trajectories (see Fig.~1), it becomes clear
that indeed, this resonance is the lowest tensor glueball.

\subsection{The analysis of the quarkonium-gluonium contents of
\newline 
$f_2(1920)$, $f_2(2020)$, $f_2(2240)$, $f_2(2300)$}

Making use of the data (\ref{7}), the expression (\ref{5}) allows us to
to find $\varphi$ as a function of the ratio $G/g$ of the coupling
constants. The result for the resonances $f_2(1920)$, $f_2(2020)$,
$f_2(2240)$, $f_2(2300)$ is shown in Fig.~4. Solid curves
enclose the values of $g_{\eta\eta}/g_{\pi^0\pi^0}$ for
$\lambda=0.6$ (this is the zone $\eta\eta$ in the $(G/g,\varphi)$
plane) and dashed curves enclose
$g_{\eta\eta'}/g_{\pi^0\pi^0}$ for $\lambda=0.6$ (the
zone $\eta\eta'$). The values of $G/g$ and $\varphi$, lying in both
zones  describe the experimental data (\ref{7}): these regions are
shadowed in Fig. 4.

The correlation curves in Fig.~4 enable us to give a qualitative
estimate for the change of the angle $\varphi$ (i.e. the relation of
the $n\bar n$ and $s\bar s$ components in the $f_2$ meson) depending on
the value of the gluonium admixture. The values $g^2$ and $G^2$ are
proportional to the probabilities of having quarkonium and gluonium
components in the $f_2$ meson, $g^2=g^2_0(1-W)$ and $G^2=G^2_0W$. Here
$W$ is the probability of a gluonium state admixture, and $g_0$ and
$G_0$ are universal constants. Since $G^2_0/g^2_0\sim1/N_c$, we take
qualitatively \begin{equation} \frac{G^2}{g^2}\ \simeq\ \frac
W{N_c(1-W)}\ . \label{9} \end{equation} Numerical calculations of the
diagrams indicate that $1/N_c$ leads to a smallness of the order of
$1/10$. Assuming that the gluonium components are less than 20\%
($W<0.2$) in each of the $q\bar q$ resonances $f_2(1920)$, $f_2(2020)$,
$f_2(2240)$, $f_2(2300)$, we have roughly $W\simeq10\,G^2/g^2$, and
obtain for the angles $\varphi$ the following intervals:
\begin{eqnarray}
&& W_{gluonium} [f_2(1920)]<20\%:
\quad-0.8^\circ<\varphi[f_2(1920)]< 3.6^\circ\ , \nonumber\\ &&
W_{gluonium}[f_2(2020)]<20\% :
\quad-7.5^\circ<\varphi[f_2(2020)]< 13.2^\circ\ , \nonumber\\ &&
W_{gluonium}[f_2(2240)]<20\%:\quad-8.3^\circ<\varphi[f_2(2240)]
<17.3^\circ\ , \nonumber\\
&&W_{gluonium}[f_2(2300)]<20\% : \quad -25.6^\circ
<\varphi[f_2(2300)] < 9.3^\circ
\label{10}
\end{eqnarray}

\section{Conclusion}

Let us summarize what we know about the status of the $(I=0, J^{PC}=
2^{++})$ mesons in the region of 1900--2400~MeV.

\begin{enumerate}

\item  The resonances $f_2(1920)$ and $f_2(2120)$ \cite{LL} (in \cite{PDG}
they are denoted as $f_2(1910)$ and $f_2(2010)$) are partners in a nonet
with $n=3$ and with a dominant $P$-component, $3\,^3P_2q\bar q$. Ignoring
the contribution of the glueball component, their flavour contents,
obtained from the reactions $p\bar p\to\pi^0\pi^0,\eta\eta,\eta\eta'$,
are
\begin{eqnarray}
f_2(1920) &=& \cos\varphi_{n=3}n\bar
n+\sin\varphi_{n=3}s\bar s, \nonumber\\ f_2(2120) &=&
-\sin\varphi_{n=3}n\bar n+\cos\varphi_{n=3}s\bar s, \nonumber\\ &&
\varphi_{n=3}\ =\ 0\pm5^\circ.
\label{13}
\end{eqnarray}
\item The next, predominantly $^3P_2$ states with $n=4$ are $f_2(2240)$
and $f_2(2410)$ \cite{LL}. (By mistake, in \cite{PDG} the resonance
$f_2(2240)$ \cite{Ani} is listed as $f_2(2300)$, while $f_2(2410)$
\cite{LL} is denoted as $f_2(2340)$). Their flavour contents at $W=0$
are determined as
\begin{eqnarray}
f_2(2240) &=&
\cos\varphi_{n=4}n\bar n+\sin\varphi_{n=4}s\bar s, \nonumber\\
f_2(2410) &=& -\sin\varphi_{n=4}n\bar n+\cos\varphi_{n=4}s\bar s,
\nonumber\\
&& \varphi_{n=4}\ =\ 5\pm11^\circ.
\label{14}
\end{eqnarray}

\item $f_2(2020)$ and $f_2(2340)$ \cite{LL} belong to the basic $F$-wave
nonet $(n=1)$ (in \cite{PDG} the $f_2(2020)$ resonance \cite{Ani} is
denoted as $f_2(2000)$ and is put in the section "Other light mesons",
while $f_2(2340)$ \cite{LL} is denoted as $f_2(2300)$). The flavour
contents of the $1\,^3F_2$ mesons are
\begin{eqnarray}
f_2(2020) &=& \cos\varphi_{n(F)=1}n\bar n+\sin\varphi_{n(F)=1}s\bar s,
\nonumber\\
f_2(2340) &=& -\sin\varphi_{n(F)=1}n\bar n+\cos\varphi_{n(F)=1}s\bar s,
\nonumber\\
&& \varphi_{n(F)=1}\ =\ 5\pm8^\circ.
\label{15}
\end{eqnarray}

\item The resonance $f_2(2300)$ has a dominant $F$-wave component; its
flavour content for $W=0$ is defined as
\begin{equation}
\label{16}
f_2(2300)=\cos\varphi_{n(F)=2}n\bar n+\sin\varphi_{n(F)=2}s\bar s,
\quad \varphi_{n(F)=2}=-8^\circ\pm12^\circ.
\end{equation}
A partner of $f_2(2300)$ in the $2\,^3F_2$ nonet has to be a $f_2$-resonance
with a mass $M\simeq2570\,$MeV. Let us stress once more that there is a
resonance, observed in the $\phi\phi$ spectrum, and denoted as $f_2(2300)$
\cite{PDG}. However, according to the re-analysis \cite{LL}, its mass
was shifted to the region $2340\pm15\,$MeV.

\item The broad $f_2(2000)$ state is the lowest tensor glueball. The
corresponding coupling constants $f_2(2000)\to\pi^0\pi^0, \eta\eta,
\eta\eta'$ satisfy the relations (\ref{6}) with $\lambda\simeq0.5-0.7$.
The admixture of the quarkonium component $(q\bar q)_{glueball}$ in
$f_2(2000)$ cannot be determined by the ratios of the coupling constants
between the hadronic channels; to define it, $f_2(2000)$ has to be
observed in $\gamma\gamma$-collisions. The value of $(q\bar q)_{glueball}$
in $f_2(2000)$ may be quite large (of the order of 50\%; indeed, the
rules of $1/N$-expansion do not forbid the mixing of $gg$ and $q\bar q$).
It is, probably, just the largeness of the quark-antiquark component in
$f_2(2000)$ which results in its suppressed production in the
radiative $J/\psi$ decays \cite{Bugg}.

We have now two observed glueballs, a scalar $f_2(1200-1600)$
\cite{APS-PL,AAS-PL} (see also \cite{book,ufn04}) and a tensor
$f_2(2000)$ one. Both these glueball states transformed into broad
resonances owing to the accumulation of widths of their neighbours. The
existence of a low-lying pseudoscalar glueball is also expected. It is
natural to assume that it should also turn into a broad resonance.
Consequently, the question is, where to look for this broad $0^{-+}$
state: it can be found both in the region of 1700~MeV, or much higher,
$\sim2300\,$MeV (see the discussion in \cite{Bugg}, Section 10.5). In
\cite{Faddeev} the idea is put forward that the lowest scalar and
pseudoscalar glueballs must have roughly equal masses. If so, a
$0^{-+}$ glueball has to occur in the 1700~MeV region.

\end{enumerate}

The authors are grateful to D.V.~Bugg, L.D.~Faddeev and S.S.~Gershtein
for stimulating discussions. The paper was supported by the grant No.
04-02-17091 of the RFFI.

\newpage

\begin{figure}
\centerline{\epsfig{file=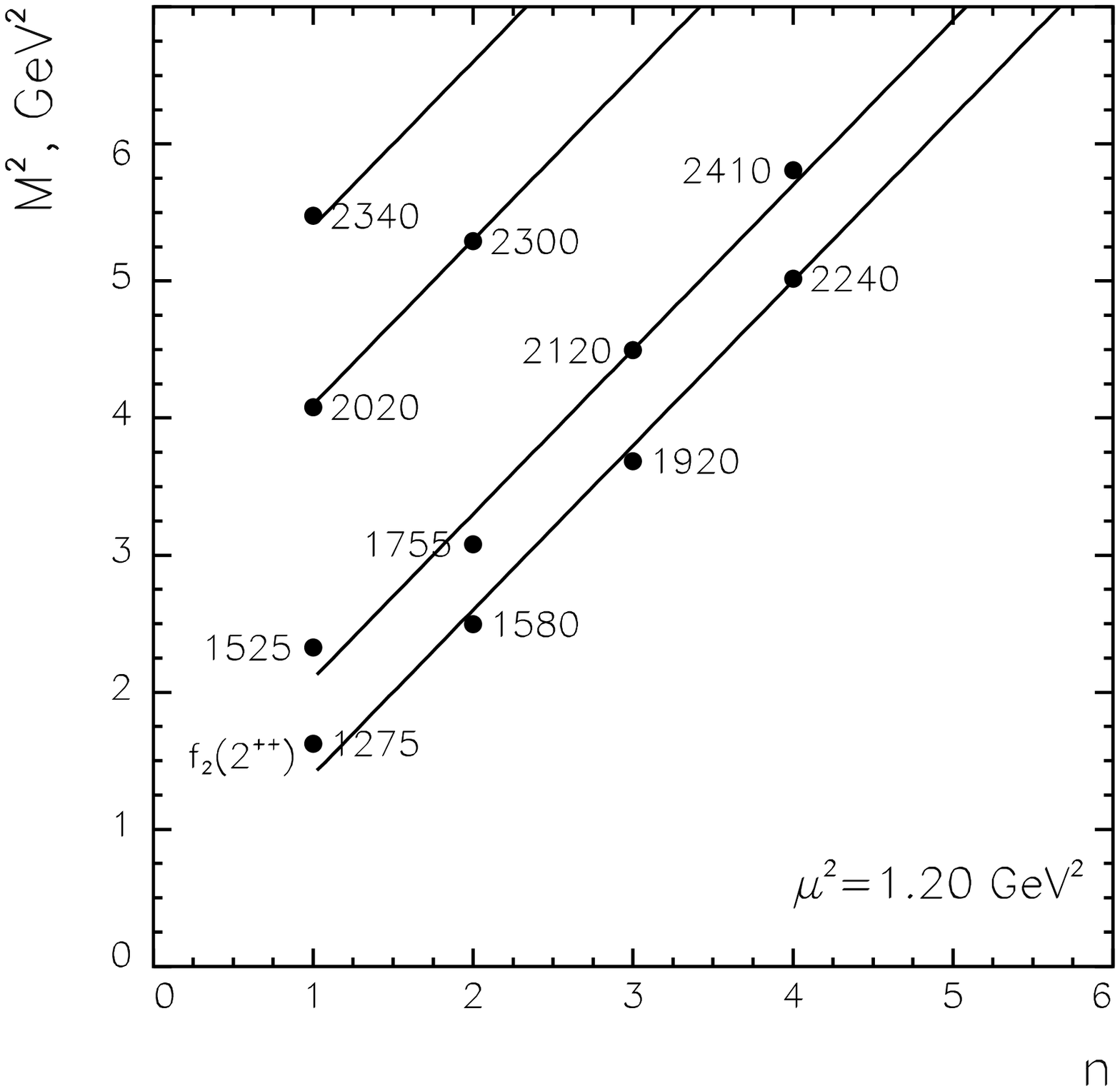,width=10cm}}
\caption{The $f_2$ trajectories on the $(n,M^2)$ plane; $n$ is the
radial quantum number of the $q\bar q$ state. The numbers stand for the
experimentally observed $f_2$-meson masses $M$.}
\end{figure}

\begin{figure}[h]
\centerline{\epsfig{file=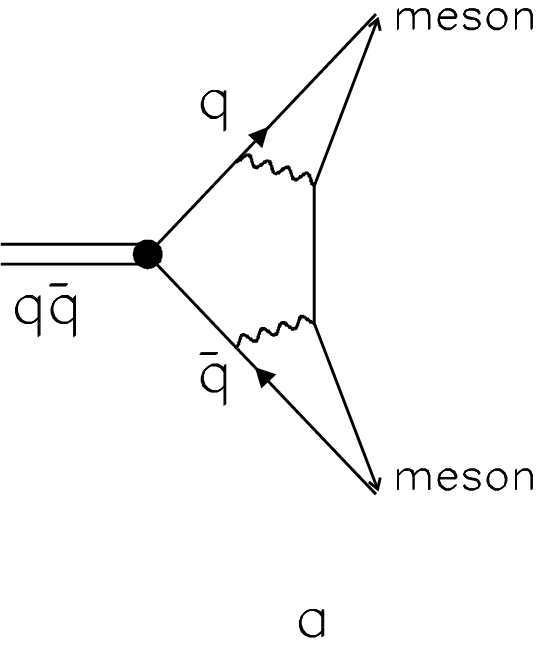,height=5cm}\hspace{0.5cm}
            \epsfig{file=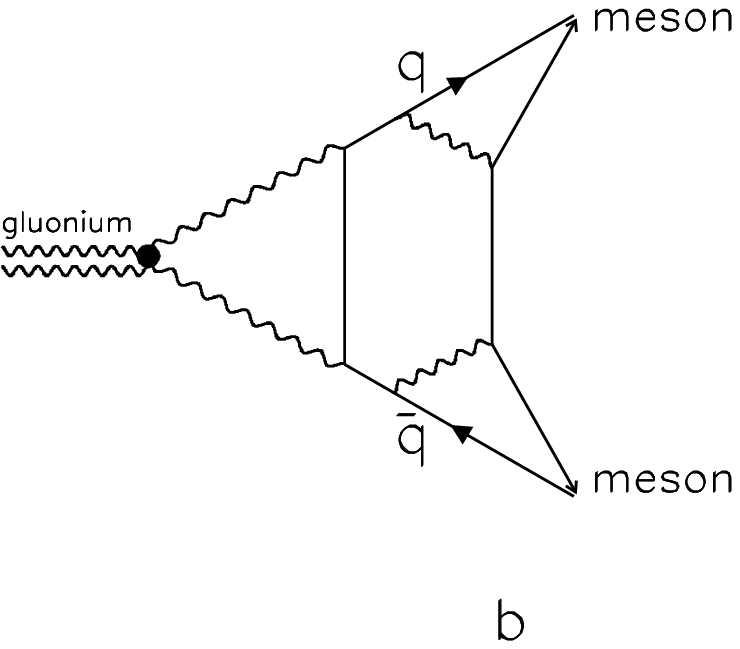,height=5cm}}
\caption{(a,b)  Examples of planar diagrams responsible for the
decay of the $q\bar q$-state and the gluonium into two $q\bar q$-mesons
(leading terms in the $1/N$ expansion).}
\end{figure}

\newpage
\begin{figure}[h]
\centerline{\epsfig{file=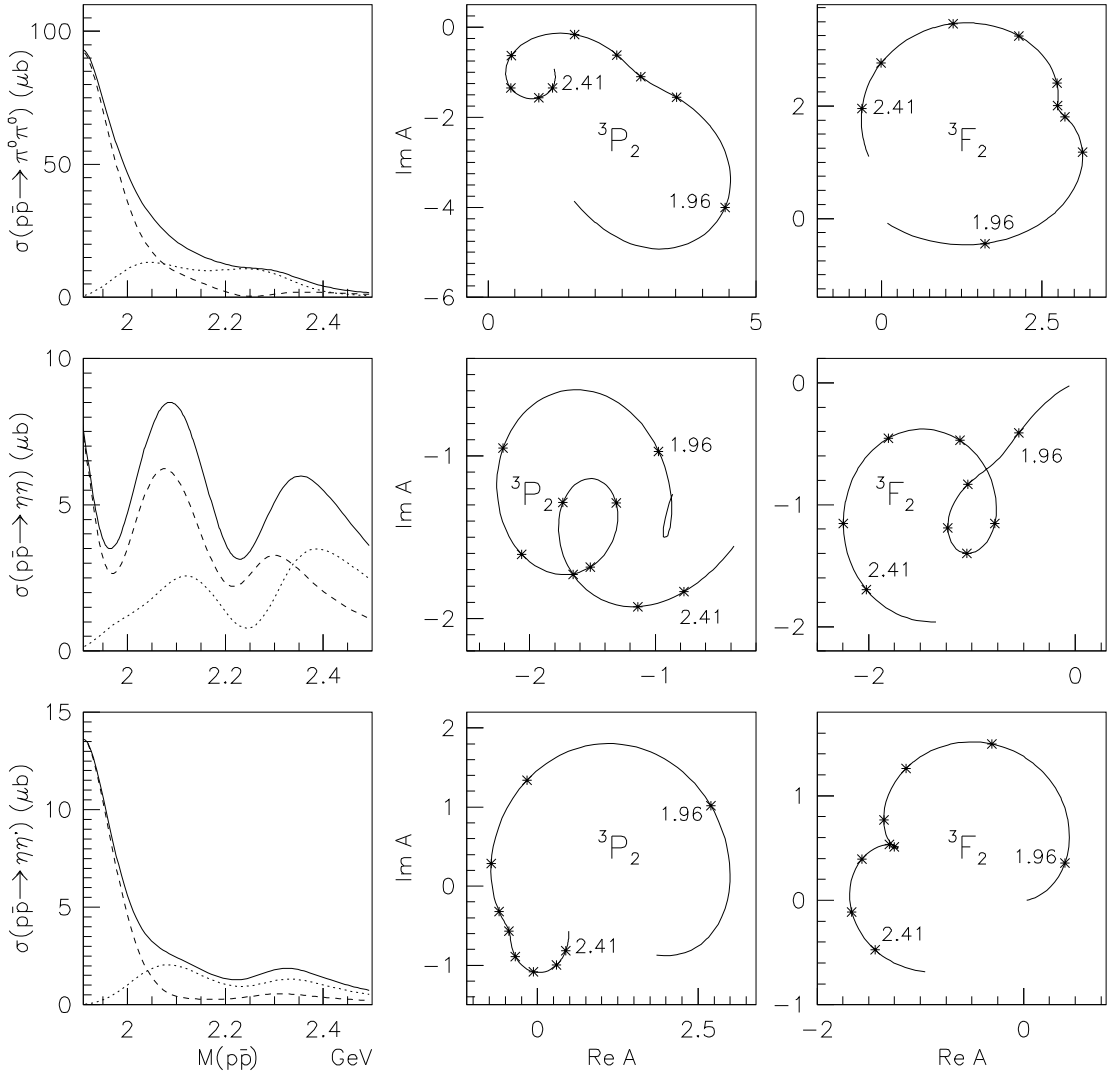,width=17cm}}
\caption{Cross sections and Argand-plots for $^3P_2$ and $^3F_2$ waves
in the reaction $p\bar p\to\pi^0\pi^0,\eta\eta,\eta\eta'$. The upper
row refers to $p\bar p\to\pi^0\pi^0$:
 we demonstrate the cross sections
 for $^3P_2$ and $^3F_2$ waves
(dashed and dotted lines, correspondingly) and the total $(J=2)$ cross
section (solid line) as well as Argand-plots for the $^3P_2$ and
$^3F_2$ wave amplitudes at invariant masses $M=1.962$, $2.050$,
$2.100$, $2.150$, $2.200$, $2.260$, $2.304$, $2.360$, $2.410$ GeV. The
figures on the second and third rows refer to the reactions $p\bar
p\to\eta\eta$  and $p\bar p\to\eta\eta'$.} \end{figure}

\begin{figure}[h]
\centerline{\epsfig{file=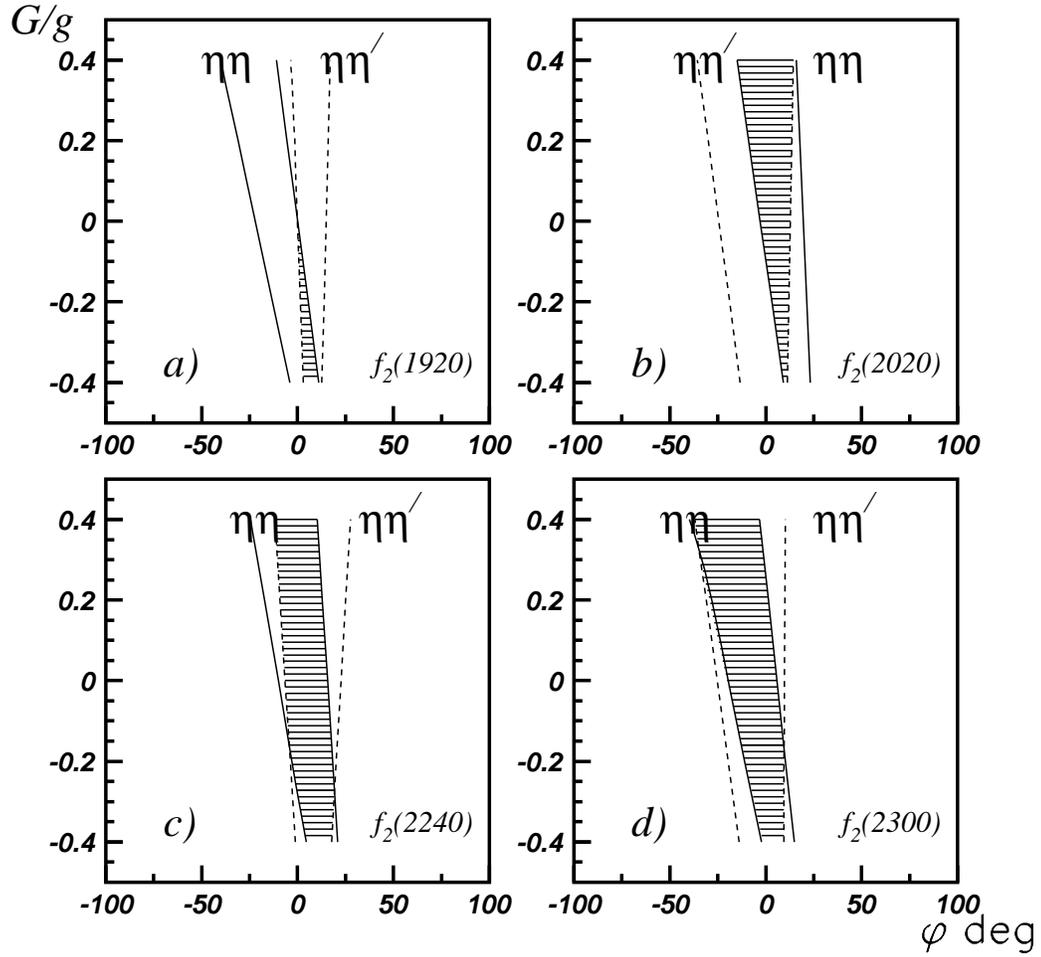,width=15cm}}
\caption{Correlation curves $g_{\eta\eta}(\varphi,G/g)
/g_{\pi^0\pi^0}(\varphi,G/g)$ and
$g_{\eta\eta'}(\varphi,G/g)/g_{\pi^0\pi^0}(\varphi,G/g)$ drawn
according to (7) at $\lambda=0.6$ for
$f_2(1920)$, $f_2(2020)$, $f_2(2240)$, $f_2(2300)$  [2,17].
  Solid and dashed curves
enclose the values
$g_{\eta\eta}(\varphi,G/g)/g_{\pi^0\pi^0}(\varphi,G/g)$  and
$g_{\eta\eta'}(\varphi,G/g)/g_{\pi^0\pi^0}(\varphi,G/g)$
which obey (9)
 (the zones $\eta\eta$ and $\eta\eta'$ in the $(G/g,\varphi)$
plane).
The values of $G/g$ and $\varphi$, lying in both
zones  describe the experimental data (9): these regions are
shadowed. }

\end{figure}

\end{document}